\documentstyle[12pt]{article}

\begin{document}

\title{\bf Effects related to spacetime foam in particle physics} 
\author{A.A. Kirillov\footnote{e-mail: kirillov@unn.ac.ru} \\ 
{\em  Institute for Applied Mathematics and Cybernetzics,}\\ 
603005 {\em Nizhnii Novgorod, Russia}} 
\date{} 
\maketitle

\begin{abstract}
It is found that the existence of spacetime foam leads to 
a situation in which the number 
of fundamental quantum bosonic fields is a variable quantity. 
The general aspects of an exact theory that allows for a 
variable number of fields are discussed,
and the simplest observable effects generated by the foam 
are estimated. It is shown that in the absence of processes 
related to variations in the topology of space, 
the concept of an effective field can be reintroduced and 
standard field theory can be restored.
However, in the complete theory the ground state is 
characterized by a nonvanishing
particle number density. From the effective-field 
standpoint, such particles are "dark".
It is assumed that they comprise dark matter of the universe.
The properties of this dark matter are discussed, 
and so is the possibility
of measuring the quantum fluctuation in the field potentials.
\end{abstract}

\section{Introduction}

In gravitation theory it is assumed that spacetime is a smooth 
manifold at scales
much greater than the Planck length, while at Planck scale 
all geometric properties disappear
and spacetime itself acquires a foamlike structure \cite{wheeler}.  
There are two basic indications 
of such behaviour of spacetime. The first is related  to the fact 
that at the Planck scale the vacuum
fluctuations of the metric and curvature are of the same order 
as the corresponding average quantities.
Not only does this follow from simple estimates - rigorous 
calculations also support
this idea. In particular, the fact that such fluctuations 
exist leads to the absence of a 
classical background space in the Planck stage of evolution 
of the early universe \cite{K97}. 
The second indication is the fact that at small scales the 
topology of space also experiences 
quantum fluctuations \cite{wheeler}. 
The study of possible observable effects related to changes 
in the topology of space is 
attracting ever more attention. In particular, to describe 
such effects, Hawking 
\cite{wormholes} used wormholes and virtual black holes. 
Another work worth noting is that of Garay
\cite{Garay},  who proposed a phenomenological method to 
account for spacetime foam.

The absence of a background space at small scales is a serious 
problem in quantum field theory. 
The possibility of resolving this problem is usually related to 
the development of nonperturbative
methods \cite{loop}, in which the concept of background fields is not used.
However, these theories also rely on the presence of a 
coordinate basis space, whose topology 
is fixed by the statement of the problem and therefore 
is not a dynamic characteristic.

This paper elaborates on a possible way to set up a quantum field 
theory in the case in which 
the topology and structure of physical space may vary. 
The main idea of this method was set forth in Ref. 
\cite{kir94} in order to describe
the quantum birth of the early universe.

The following observation forms the basis of the proposed method.
On the one hand, as noted earlier, variations in the topology of space can occur 
at scales where the very concept of a smooth
manifold breaks down, at least due to the presence of vacuum 
fluctuations. 
On the other hand, it is believed that there is no other way 
to describe the given 
region but to extrapolate the spatial relationships existing 
at larger scales to it.
In other words, all possible topologies of physical space should be 
described in terms of a
consistent coordinate basis space. We call this space simply a basis.
Since measurement instruments, which are classical objects, 
play a fundamental role
in quantum theory \cite{Landau}, 
it is expected that the properties of the basis are determined entirely
by the measuring device.

If we specify the quantum state corresponding to a fixed 
topology of physical space and if the topology differs 
substantially from that of the basis,
the image of physical space in terms of the basis coordinates
cannot be one-to-one. In the same way, when functions defined 
in physical space and corresponding to different physical 
observables are mapped to the basis space, they cease
to be single-valued and become multivalued functions of 
the coordinates.
Furthermore, the number of images of an arbitrary physical
 observable is an 
additional variable quantity, which generally speaking, 
depends
on the position in the basis space.

Thus, we arrive at a situation in which the number of  
fields corresponding to a physical observable is a variable 
quantity. In quantum theory this variable is an operator
whose eigenvalues characterize the topological structure of
space. The possible dependence of this quantity on spatial
coordinates means that the given quantity is a characteristic
or measure of the number density of the degrees of freedom
of the field.

A natural way to describe systems with a variable number
of degrees of freedom  is to use second quantization. Before
we begin to describe the method as applied to the problem
in question, we make the following remark. In the 
standard second-quantization method, the number of degrees
of freedom characterizes the number of particles or elementary
excitations (quanta) in the system. Here it is assumed
that the particles obey the identity principle or, as it is said,
the indistinguishability principle. It can be expected that in
measurements at small scales the different images of the 
same physical observable also obey the identity principle\footnote{
Thus, we assume the existence of a fundamental restriction on the degree 
of accuracy for measurements of different field observables. In quantum 
gravity this is well- known restriction, which represents the existence of 
a minimal quantum uncertainty for metric of the type 
$\Delta g \sim L_{pl }^2 / L^2 $, 
where $L_{pl}$ is the Planck length and $L$ is the characteristic size
of the region in which the measurement takes place. We note that 
such restrictions are not something very unusuall
in quantum mechanics. It suffices to recall the impossibility to detect
a position of a relativistic particle with the accuracy exceeding 
the Compton length, which seems to be  in a close analogy with the 
restriction pointed out above.}.
Indeed, the possibility of distinguishing between the different
images of observables would mean that physical space in
itself has certain topology and structure, which by assumption 
is impossible (at least in view of the presence of quantum fluctuations
in the topology).

Two types of statistics, Bose and Fermi, exist for particles,
depending on the symmetry of the wave function under
particle permutations. Accordingly, we must also select
the type of statistics when performing second quantization of
the degrees of freedom of the fields. Since second quantization
reflects the properties and topology of physical space,
this selection must be unique for all types of fields and physical
quantities. Here it turns out that the only acceptable
choice is Fermi-Dirac statistics, since otherwise in dealing
with fermions we immediately confront a violation of the
Pauli principle.

At a fundamental level, the composition of matter is determined
by a set of fields and their sources. The sources are
point particles, which in quantum theory behave like fermions.
The need to perform second quantization of the sources 
arises already in relativistic theory and hence no changes in
the description of fermions emerge. A new interpretation is
added, however. For instance, pair production corresponds to
a change in the structure of physical space (it can be said that
processes related to changes in the properties of space proceed
much more easily at isolated points than they do in
entire regions).

When fields are quantized, the idea of particles, the
quanta of a field, also emerges. Such particles, however,
obey the Bose-Einstein statistics. Here, generally speaking,
particle production is not associated with variations in the 
topology of space. There is a certain similarity between this
aspect and the situation in solid state physics, where excitation
of vibrations in a crystal lattice (phonon production) is
not associated with variations in the true number of degrees
of freedom.

Thus, the variation mentioned above primarily involves
bosonic fields.

\section{General scheme of second quantization of fields}

We consider a set  $M$, which in the future acts like a basis 
manifold, and specify an arbitrary field  $\varphi $ on it. We also
assume that there is a device that can do complete measurements
of the quantum states of the field. A complete measurement
can always be expanded in a set of elementary
measurements. For instance, to make a complete measurement 
of a field state we must measure the field amplitude at 
every point $x\in M$, or equivalently measure the number of
particles (or amplitude) in each Fourier mode. Thus, the
device can be viewed as a set of elementary detectors.

Let  $A$ be the set of possible readings of an elementary
detector\footnote{The set  $A$ can be called an elementary system 
of quantum numbers.}. 
The structure of   $A$ can be described in the following
way.  In  $A$ we select an arbitrary system of coordinates  $\xi $. 
Generally, there is a natural projection operator  $P$ ($P^{2}=P$) 
that partitions the coordinates into two groups: 
$\xi =\left( \left( I-P\right) \xi ,P\xi \right) =$ 
$\left( \eta ,\zeta \right) $, where  $I$ denotes the identity operator.
The first group,   $\zeta $, refers to the manifold  $M$ and describes the
position in space  $M^{\ast }$ at which the elementary measurement 
takes place (here and in what follows $M^{\ast }$ denotes either
space $M$ or the mode space). The second group, $\eta $, refers to
the field  $\varphi $ and describes the position in space  $V$. The coordinate
 $\eta \in V$  denotes either the field amplitude or the number
of particles corresponding to the field. Thus, the set $A$ acquires
the features of a fiber space with basis $P\left( A\right) \sim M^{\ast }$  and 
fiber $P^{-1}\left( \zeta \right) =V$. The result of a complete measurement of
field  $\varphi $ is a fibration section, which is the map  
$\varphi  : M^{\ast }\rightarrow A$. 
What is important is that in the usual picture an arbitrary
section intersects each fiber only once, i.e., the projection of
the section coincides with the space  $M^{\ast }$  
($P\left( \varphi \right) \equiv M^{\ast }$), which
implies that such sections can be represented by functions
 $\eta \left( \zeta \right) $ on  $M^{\ast }$ with values in  $V$.

As noted above, the topology and geometric structure of
the set  $A$ (and thus of  $M^{\ast }$) reflects the macroscopic properties
of the measurement process.  On the other hand, the real
physical space $M_{ph}$ is assumed to have arbitrary topology and 
structure\footnote{When speaking of the topology of the physical space,
we mean either the topology of the space $M_{ph}$ itself, or the topology 
of the related space 
$M_{ph}^{\ast }$,  depending on the quantities being measured. Note, however,
that the relationship between these two spaces is nontrivial.}.
Furthermore, in a general quantum state, the
properties of space  $M_{ph}^{\ast }$ are, generally speaking, not fixed.
Thus, a physical field must be defined as an extended section
of the form $\widetilde{\varphi }:M_{ph}^{\ast }\rightarrow A$. 
Here an arbitrary section can intersect
each fiber an arbitrary number of times. Furthermore, if
the topology of space $M_{ph}^{\ast }$ changes, so does the number of
intersections. Thus, the number of images of field  $\widetilde{\varphi }$ in space
 $M^{\ast }$ is variable. An image of space  $M_{ph}^{\ast }$ is a subset in
 $M^{\ast }$ ($P\left( \widetilde{\varphi }\right) =M_{ph}^{\prime }\subset M^{\ast }$)
that can be represented as a union of
distinct pieces, $M_{ph}^{\prime }=\bigcup \sigma _{j}$, so that on each piece
$\sigma _{j}$ the field  is described by a given number of functions  
$\eta _{i}\left( \zeta \right) $, 
$\zeta \in \sigma _{j}$ ($i=1,2,\ldots m,$ where $m$  is an 
integer characterizing the
number of images of the space  $M_{ph}^{\ast }$ in  $\sigma _{j}$).  
Note that in general
the dimensionality of the pieces $\sigma _{j}$ can differ from the 
dimensionality of  $M^{\ast }$.

Thus, if the topology of physical space is an additional
degree of freedom, the result of a complete measurement of
the state of the field will be represented by a definite set of
functions  $\left\{ \eta _{J}\left( \zeta 
\right) \right\} $, ($J=\left( i,\sigma \right) $ and $\zeta \in \sigma $).  
Formally, such states can be classified in the following way.

We introduce a set of operators $C^{+}\left( \xi \right) $  
and $C\left( \xi \right) $, the 
creation   and annihilation operators for an individual element
of the set  $A$. For the sake of simplicity we assume that
the measure of each individual point $\xi \in A$ is finite (as in the
case in which the coordinates $\xi $ take discrete values). We 
require that these operators satisfy the anticommutation relations
\begin{equation} 
\left\{ C\left( \xi \right) C^{+}\left( \xi ^{\prime }\right) \right\} =C\left( 
\xi \right) C^{+}\left( \xi ^{\prime }\right) +C^{+}\left( \xi ^{\prime 
}\right) C\left( \xi \right) =\delta _{\xi \xi ^{\prime }}.  \label{CCR} 
\end{equation} 
We define the vacuum state $\left| 0\right\rangle $ by the relationship
$C\left( \xi \right) \left| 0\right\rangle =0$ 
and build a Fock space $F$ in which the basis
consists of the vectors  ($n=1,2,\ldots $) 
\begin{equation} \left| \xi _{1},\xi _{2},\ldots ,\xi _{n}\right\rangle 
=\prod_{i=1}^{n}C^{+}\left( \xi _{i}\right) \left| 0\right\rangle .  
\label{FSTS} 
\end{equation} 
The vacuum state corresponds to complete absence of a field
and hence of the observables associated with the field. The
state $\left| \xi \right\rangle $ describes the field $\varphi $ 
with only one degree of freedom.
This can be either a field concentrated at a single point
or a field containing only one mode, and the quantity $\xi \in A$ 
describes the intensity (the number of quanta) and the position
of the field in  $M^{\ast }$. States described by single-valued
functions are constructed in the following way:
 \begin{equation} 
\left| \eta \left( \zeta \right) \right\rangle =\prod_{\zeta \in M^{\ast 
}}C^{+}\left( \eta \left( \zeta \right) ,\zeta \right) \left| 0\right\rangle ,  
\label{SS} 
\end{equation} 
where the direct product is taken over the entire space $M^{\ast }$ ,
and where we have partitioned the coordinates  $\xi $ into two
groups: $\xi =$ $\left( \eta ,\zeta \right) $. Generally, such states do not belong to a
Fock space. Furthermore, when the coordinates $\zeta \in M^{\ast }$ run
through continuous values, this expression requires an extension
of its definition and hence can be interpreted only formally.
However, when the variations of the physical quantities
in real processes involve only a finite part of the set $M^{\ast }$,
we can stay within a Fock space.

We now examine an arbitrary domain $\sigma \in M^{\ast }$ and
define a set  of operators
\begin{equation}
D^{+}\left( \eta \left( \zeta \right) ,\sigma \right) =\prod_{\zeta \in
\sigma }C^{+}\left( \eta \left( \zeta \right) ,\zeta \right) ,
\end{equation}
where  the domain of the function $\eta \left( \zeta \right) $  is limited to the set
 $\sigma $.  Then the states with an arbitrary number of fields can be written
 \begin{equation} \left| \eta _{1},\eta _{2}\cdots ,\eta 
_{n}\right\rangle =\prod_{i=1}^{n}D^{+}\left( \eta _{i}\left( \zeta \right) 
,\sigma _{i}\right) \left| 0\right\rangle .  \label{sts} \end{equation} 
The interpretation of these states is obvious. Suppose that all
functions $\eta _{i}\left( \zeta \right) $ are specified on a single set  $\sigma $.
Then in the given domain a complete measurement will show the presence
of a set consisting of $n$ different fields  $\eta _{1}\left( \zeta \right) $, $\eta _{2}\left( \zeta \right) $, ..., $\eta 
_{n}\left( \zeta \right) $.  It is convenient to introduce the number density
operator of the fields:
\begin{equation} N\left( \zeta \right) =\sum_{\eta \in 
V}C^{+}\left( \eta ,\zeta \right) C\left( \eta ,\zeta \right) .  \label{N1} 
\end{equation} 
Then for  $\zeta \in \sigma $ the states (\ref{sts}) represent the eigenstates of the 
operator $N\left( \zeta \right) $ with  eigenvalues
 \begin{equation} N\left( \zeta \right) \left| \eta 
_{1},\eta _{2}\cdots ,\eta _{n}\right\rangle =n\left| \eta _{1},\eta _{2}\cdots 
,\eta _{n}\right\rangle .  \end{equation}
Clearly, the states with a fixed number of fields correspond
to a fixed topology of the space $M_{ph}^{\ast }$.  Then under certain
conditions (the requirement  that the functions  $\eta _{i}\left( \zeta \right) $ be 
smooth at cuts), instead of the set of functions $\eta _{i}\left( \zeta \right) $ we can
introduce a single-valued function  $\eta \left( \zeta \right) $  and thus restore the
structure of the set $M_{ph}^{\ast }$. Conversely, each space $M_{ph}^{\ast }$ can be
projected on the basis  $M^{\ast }$ by performing the necessary
paste-up,  so that the state vector of the field takes the form (\ref{sts}).

The space $H$ formed by the vectors  (\ref{sts}) and their superposition
lays the basis for building the Hilbert space of the 
theory. An arbitrary operator  $\widehat{O}\left( \xi \right) $ related to the field (and
symmetrized in the number of fields) can be expressed in
the standard way in the terms of the set of basis operators $C$ 
and $C^{+}$ : 
$$\widehat{O}=\sum  D_{I}^{+}O_{IJ}D_{J}$$
(where  $I,J=\left( \eta _{i}\left( \zeta \right) ,\sigma 
\right) $, and   $\sigma $ is an arbitrary domain in $M^{\ast }$), thus
defining the action of this operator in $H$. The
specific way in which this Hilbert space is built is determined
by the physical problem at hand.

\section{Scalar field in the second - quantization representation}

In Sec. 2 we discussed the general scheme of second
quantization, irrespective of the dynamics of the field. We
now turn to the example  of a real scalar field  $\varphi $ (the generalization
to the case of arbitrary fields is obvious).  For the 
basis space we take ordinary flat Minkowski space.

One idea that is central to particle physics is the representation
in which quantum states of a field are classified in
terms of physical particles. Since quantum states of a field
can in general contain an arbitrary number of identical
modes, the definition of particles and their relation to field
operators require certain modifications. We find it more convenient 
to operate with discrete indices. To this end we 
require that the field in question be located in a cube with edge
length $L$, and we introduce periodic boundary conditions. As
necessary, we can replace sums with integrals (as $L\rightarrow \infty $ ) via
the usual prescription: $\sum \rightarrow \int \left( L/2\pi \right) ^{3}d^{3}k$.

We now examine the expansion of the field operator $\varphi $ 
in plane waves,
\begin{equation} \varphi \left( x\right) =\sum_{k}\left( 2\omega 
_{k}L^{3}\right) ^{-1/2}\left( a_{k}e^{ikx}+a_{k}^{+}e{\ }^{-ikx}\right) ,  
\label{field} \end{equation} 
where  $\omega _{k}=\sqrt{k^{2}+m^{2}} $, and $k=2\pi n/L$, with $n=\left( n_{x},n_{y},n_{z}\right) $. 
The general expression for the Hamiltonian is
\begin{equation} H=H_{0}+V  \label{Hf} \end{equation} 
where $H_{0}$ describes free particles, 
\begin{equation} H_{0}=\sum_{k}\omega 
_{k}a_{k}^{+}a_{k}\,\,+e_{k},  \label{H}
 \end{equation} 
and the potential term $V$ is responsible for the interaction,
and can be represented in the normal form:
\begin{equation} 
V=\sum_{n,\left\{ m\right\} ,\left\{ m^{\prime }
\right\} }
V_{\left\{ m\right\} ,\left\{ m^{\prime }\right\} }^{n} 
\end{equation} \begin{equation} V^{n}{}_{\left\{ m\right\} ,\left\{ m^{\prime 
}\right\} }=\sum_{k_{1},\ldots k_{n}}^{\prime }V_{\left\{ m\right\} ,\left\{ 
m^{\prime }\right\} }^{n}\left( k_{1},\ldots k_{n}\right) \prod_{i=1}^{n}\left( 
a_{k_{i}}^{+}\right) ^{m_{i}}\left( a_{k_{i}}\right) ^{m_{i}^{\prime }}, 
\label{Int} \end{equation} 
Here we assume that the sum with respect to the wave vectors 
$k_{i}$  contains no terms with equal indices, i.e., $k_{i}\neq k_{j}$ for
any pair of indices $i$ and  $j$ (the sum is taken over distinct 
modes), and allow for the fact  that for different wave numbers
the operators $a_{k_{i}}$ and $a_{k_{j}}^{+}$ commute.

The quantity  $e_{k}$ in  (\ref {H}) is the energy of the ground state
of the $k$th mode. In a flat space without particles, the energy
must be zero, so we assume that $e_{k}=0$ throughout the
present paper. However, as we show in the sections that follow, 
the nontrivial  nature of the topology of the space generally 
leads to a value of  $e_{k}$ that is finite. Note that the dependence
of the zero energy on the topology of space is known
as the Casimir effect  \cite{Cas48}  and is assumed to be an experimentally
established fact \cite{Sp58}.

When the number of modes is variable, the set of field
operators  $\left\{ a_{k},a_{k}^{+}\right\} $ is replaced by the somewhat expanded set
 $\left\{ a_{k}\left( j\right) ,a_{k}^{+}\left( j\right) \right\} $, where 
$j\in \left[ 1,\ldots N_{k}\right] $, and   $N_{k}$ is the number
of modes for a given wave number  $k$. For a free field the
energy is an additive quantity, which can be written 
\begin{equation} H_{0}=\sum_{k}\sum_{j=1}^{N_{k}}\omega _{k}a_{k}^{+}\left( 
j\right) a_{k}\left( j\right) \,\,.  \end{equation} 
Since the modes are indistinguishable,  the interaction operator 
has the obvious generalization 
\begin{equation} 
V_{\left\{ m\right\} ,\left\{ m^{\prime }\right\} }^{n}=\sum_{k_{1},\ldots 
k_{n}}^{\prime }\sum_{j_{1},\ldots j_{n},}V_{\left\{ m\right\} ,\left\{ 
m^{\prime }\right\} }^{n}\left( k_{1},\ldots k_{n}\right) \prod_{i=1}^{n}\left( 
a_{k_{i}}^{+}\left( j_{i}\right) \right) ^{m_{i}}\left( a_{k_{i}}\left( 
j_{i}\right) \right) ^{m_{i}^{\prime }}, \end{equation}
where the indices  $j_{i}$  run through the corresponding intervals 
$j_{i}\in \left[ 1,\ldots N\left(  k_{i}\right) \right] $. 
It is convenient to introduce the notation
\begin{equation} 
A_{m,n}\left( k\right) =\sum_{j=1}^{N\left( k\right) } \left( a_{k}^{+}\left( 
j\right) \right) ^{m}\left( a_{k}\left( j\right) \right) ^{n} .  \label{A2} 
\end{equation} 
Then the expression for the field Hamiltonian takes the form
\begin{equation} H=\sum_{k}\omega _{k}A_{1,1}\left( k\right) +\sum_{n,\left\{ 
m\right\} ,\left\{ m^{\prime }\right\} }\sum_{k_{1},\ldots k_{n}}^{\prime 
}V_{\left\{ m\right\} ,\left\{ m^{\prime }\right\} }\left( k_{1},\ldots 
k_{n}\right) \prod_{i=1}^{n}A_{m_{i},m_{i}^{\prime }}\left( k_{i}\right) ,  
\label{Ham} \end{equation}

We can now express the main quantities in terms of the 
fundamental operators $C^{+}\left( \xi \right) $ and $C\left( \xi \right) $.
For the operators $a$
and $a^{+}$ it is convenient to use the Fock - Bargmann representation,
in which  operators act in the space of entire analytic
functions with a scalar product of the type
\begin{equation} \left( f,g\right) =\int f^{\ast }\left( 
a\right) g\left( a^{\ast }\right) e^{-a^{\ast }a}\frac{da^{\ast }da}{2\pi i}, 
\end{equation}
the action of these operators is defined as
\begin{equation} a^{+}f\left( 
a^{\ast }\right) =a^{\ast }f\left( a^{\ast }\right) ;\;\;af\left( a^{\ast 
}\right) =\frac{d}{da^{\ast }}f\left( a^{\ast }\right) .  
\end{equation} 
Then for the normal field coordinates we can take the
complex - valued quantities $a^{\ast }$; thus, the set $A$ consists of the
pairs $\xi =\left( a^{\ast },k\right) $. For the fundamental operators $C^{+}\left( \xi \right) $ and
 $C\left( \xi \right) $  it is convenient to use the representation 
\begin{equation} C\left( a^{\ast },k\right) =\sum_{n=0}^{\infty }C\left( 
n,k\right) \frac{\left( a^{\ast }\right) ^{n}}{\sqrt{n!}},\;\;C^{+}\left( 
a,k\right) =\sum_{n=0}^{\infty }C^{+}\left( n,k\right) \frac{a^{n}}{\sqrt{n!}}.  
\label{C} \end{equation} 
Then the anticommutation relations (\ref{CCR}) become 
\begin{equation} \left\{ C\left( n,k\right) ,C^{+}\left( m,k^{\prime 
}\right) \right\} =\delta _{n,m}\delta _{k,k^{\prime }}\,\,.  \label{3.3m} 
\end{equation} 
The physical meaning of the operators $C\left( n,k\right) $ and  $C^{+}\left( n,k\right) $ 
is that they create and annihilate modes with a given number
of particles.

Now, to  express the Hamiltonian (\ref{Ham}) in terms of
$C\left( n,k\right) $ and  $C^{+}\left( n,k\right) $ it suffices to derive  the corresponding
expressions for the operators  (\ref{A2}). In the second - quantization
representation, the expressions  for the given operators are
defined to be
\begin{equation} 
\widehat{A}_{m,n}\left( k\right) =\int e^{-a^{\ast }a}\frac{da^{\ast }da}{2\pi 
i} C^{+}\left( a,k\right) \left( a^{\ast }\right) ^{m}\left( \frac{d}{da^{\ast 
}}\right) ^{n}C \left( a^{\ast },k\right)  \label{A1} \end{equation}
or,  with allowance for (\ref{C}),
\begin{equation} 
\widehat{A}_{m _{1}, m _{2}}\left( 
k\right) =\sum\limits_{n=0}^{\infty }\frac{\sqrt{\left( n+m_{1}\right) !\left( 
n+m_{2}\right) !}}{n!}C^{+}\left( n+m_{1},k\right) C\left( n+m_{2},k\right) .  
\label{A} \end{equation}
An expression for the Hamiltonian in terms of the operators 
$C^{+}\left( \xi \right) $ and  $C\left( \xi \right) $ can
be obtained by simply substituting  (\ref{A}) into  (\ref{Ham}).
For a free field, the eigenvalues of the Hamiltonian
take the form
\begin{equation} 
\widehat{H}_{0}=\sum_{k}\omega _{k}\widehat{A}_{1,1}\left( k\right) 
=\sum_{n,k}n\omega _{k}N_{n,k}\,,  \label{zu} \end{equation} 
where  $N_{n,k}$  is the number of modes for fixed values of the 
wave number $k$ and the number of particles $n$
($N_{n,k}=C^{+}\left( n,k\right) C\left( n,k\right) $).

Thus, the field state vector  $\Phi $ is a function of the occupation
numbers $\Phi \left( N_{k,n},t\right) $,   and its evolution is described by
the Schr\"odinger equation
\begin{equation} i\partial _{t}\Phi =H\Phi .  
\end{equation} 
Consider the operator 
\begin{equation} N_{k}=\sum_{n=0}^{\infty 
}C^{+}\left( n,k\right) C\left( n,k\right) .  \label{N} 
\end{equation} 
Physically, this operator characterizes the total number of
modes for a fixed wave number $k$. One can easily verify that
for the Hamiltonian  (\ref{Ham}),   $N_{k}$ is a constant of the motion,
\begin{equation} \left[ N_{k}, H\right] =0 
\end{equation}
and in this way Hamiltonians like (\ref{Ham}) preserve the topological
structure of the field. In the course of evolution, the number
of modes for each $k$ does not change.

We now turn  to the problem of representing the particle
creation and annihilation operators in this formalism. Since
the individuality of the modes is limited, operators of  type 
(\ref{A}) act like the set of operators 
$\left\{ a_{k}\left( j\right) ,a_{k}^{+}\left( j\right) \right\} $.
Among the operators  (\ref{A}) are some that change the number of particles 
by one:
\begin{equation} b_{m}\left( 
k\right) =\widehat{A}_{m,m+1}\left( k\right) ,\;\;b_{m}^{+}\left( k\right) 
=\widehat{A}_{m+1,m}\left( k\right) , \label{b} 
\end{equation} 
\begin{equation} 
\left[ \widehat{n},b_{m}^{(+)}\left( k\right) \right] =\pm b_{m}^{(+)}\left( 
k\right) ,\;\;\;\left[ H_{0},b_{m}^{(+)}\left( k\right) \right] =\pm \omega 
_{k}b_{m}^{(+)}\left( k\right) , \end{equation} 
where
\begin{equation} 
\widehat{n}=\sum_{k}\widehat{n}_{k}=\sum_{n,k}nN_{n,k}.  \label{Nk} 
\end{equation} 
Then the ground state  $\Phi _{0}$  of the field can be  defined as a
vector satisfying the relationships  ($\ m=0,1,\ldots $) 
\begin{equation} 
b_{m}\left( k\right) \Phi _{0}=0  \label{0}
 \end{equation}
and corresponding  to the minimum energy for a fixed mode
distribution $N_{k}$.  Note that in contrast to standard theory, the
ground state is generally characterized by a nonvanishing
particle number density $\widehat{n}\Phi _{0}=n_{0}\Phi  _{0}$. 
Using the vector  $\Phi _{0}$,
we can build a Fock space $F$ whose basis consists of vectors
obtained by cyclic application of the operators $b_{m}^{+}\left( k\right) $ 
to  $\Phi _{0}$.

\section{Effective field}

In the absence of processes related to changes in the 
topology of space and for a mode distribution of the form 
$N_{k}=1$ (there is only one mode for each wave number $k$), the
standard field theory is restored. Furthermore, there is a
fairly general case in which the concept of an effective field
can be introduced to restore the standard picture.

Indeed, consider the case in which the interaction operator
in   (\ref{Ham}) is expressed solely in terms of the set of operators
$b_{0}\left( k\right) $ and $b_{0}^{+}\left( k\right) $. 
Them instead of the complete Fock space $F$
we can limit ourselves to its subspace $F^{\prime }\subset F$ formed by the
cyclic application of the operators $b_{0}^{+}\left( k\right) $ to the field ground
state  $\Phi _{0}$.  If the initial state vector $\Phi $  
belongs to $F^{\prime }$, then as
the system evolves,  $\Phi \left( t\right) \in F^{\prime }$ for all  $t$ 
(at least as long as the
number of particles created remains finite).

We define the operators
\begin{equation} 
a_{k}=N_{k}^{-1/2}b_{0}\left( k\right) 
,\;\;\;a_{k}^{+}=N_{k}^{-1/2}b_{0}^{+}\left( k\right) ,  \label{norm} 
\end{equation} 
where $N_{k}$ is the operator defined in  (\ref{N}), which, when restricted
to the Fock space $F^{\prime }$,  is an ordinary number function.
For  (\ref{3.3m}) and  (\ref{A}) we find that the commutation 
relations for  $a_{k}$ and $a_{k}^{+}$ have the standard form
\begin{equation} \left[ 
a_{k},a_{k^{\prime }}^{+}\right] =\delta _{k k^{\prime }}.  \end{equation}

Thus, if the basic observable objects are particles, it is
possible to revert to the usual picture in which the particles
are quanta of an effective field $\widetilde{\varphi }$ of  type (\ref{field}). 
Note that if the
field potentials  $\varphi \left( x\right) $ are measurable quantities, then the true
expression for the field operators has the same form  (\ref{field}),
where instead of the operators $a_{k}$ and 
$a_{k}^{+}$ we must put $b_{0}\left( k\right) $
and $b_{0}^{+}\left( k\right) .$
The expression for the effective-field energy operator
has the form (\ref{Hf}), but the ground-state energy in the $k$th
mode,  $e_{k}$, must be assumed not to vanish. The value of this
energy can be found in the complete theory.

Since the operators  $a_{k}$ and $a_{k}^{+}$ reflect only some of the 
information about the state of the system, the auxiliary nature
of the effective field becomes manifest. Indeed, the only
observables related to the effective field are those particles
that outnumber the particles in
the ground state. For the particle number operator in  $F^{\prime }$  we have
\begin{equation} a_{k}^{+}a_{k}=\delta 
\widehat{n}_{k}=\widehat{n}_{k}-\overline{n}_{k}, 
\end{equation} 
where $\widehat{n}_{k}$ is the operator defined in (\ref{Nk}), 
and  $\overline{n}_{k}$
can be found by solving  $\widehat{n}_{k}\Phi _{0}=\overline{n}_{k}\Phi _{0}$.
Thus, the properties of the ground state $\Phi _{0}$ remain
beyond the scope of the effective field.

\section{Properties of the field ground state}

Equations (\ref{3.3m}) and  (\ref{A})  imply that a true vacuum state 
has the property that all field modes (and hence all observables
related to the field) are absent. A true vacuum state is
one in which there are no particles and no zero-point oscillations
related to particles. This situation is similar to the situation
in solid state physics, where in the absence of a crystal
there can be no phonons and no zero-point lattice vibrations.
Since the properties of physical space  are determined by the
properties of material fields, we conclude that in a true
vacuum state there can be no physical space. Obviously, in
reality such a state cannot be achieved.

At first glance the most common situation in particle
theory is the one in which physical space is ordinary flat 
Minkowski  space, and nontrivial topology is manifest at the
Planck scale (this is the conventional view; see Refs. 
\cite{wheeler} and \cite{Garay}). 
But since operating at the Planck scale requires using
energies unattainable with present-day accelerators, and also
requires serious consideration of quantum gravity effects, it
would appear to be impossible to make any sort of directly
measurable predictions with this theory.

In reality, the situation may be somewhat different. First, the 
stability of the Minkowski space means that probably even at
the Planck scale the topology of the space can be assumed to
be  simple (i.e.,  $N_{k}=1$ and as $k\geq k_{pl}$), at 
least as long as we do
not consider processes in which real particles with Planck
energies are produced (naturally, virtual processes cannot
lead to real changes in the topology of space).

Second, recall that the universe has already passed the
quantum stage, in which real processes involving  changes in
spatial topology might occur. After the quantum stage, processes
with topology variations are suppresed, and we can
say that the topological structure of space  has been "tempered",
so that the structure of the space is preserved as the
universe expands. Thus, we expect that at the present time
the nontrivial topology of space is most likely manifested on a
cosmological scale.

In the foregoing theory, the structure of space is determined
by the number density of the field modes. These
modes are in turn governed by Fermi statistics, i.e., they act
like a Fermi gas. To simplify matters, we examine free fields,
since consistent allowance for the interaction of field
warrants a separate investigation. We assume that the field-mode 
distribution was thermal in the Planck period of the evolution
of the universe. As  the universe expands, the temperature
drops and the gas becomes degenerate, with the field winding
up in the ground state. Thus, the field ground state $\Phi _{0}$
can be characterized by occupation numbers of the type
\begin{equation} N_{ 
k,n}=\theta \left( \mu _{k}-n\omega _{k}\right) ,  \label{GST} 
\end{equation} 
where $\theta \left( x\right) $ is the Heaviside step function and 
$\mu _{k}$ is the
chemical potential. Note that when the expansion is adiabatic,
we must put $\mu _{k}=\mu .$  When the evolution of the universe
includes an inflationary period \cite{inf,linde90},  the adiabaticity 
condition can be violated, which generally leads to additional
dependence of the chemical potential on the wave number.
For the mode spectral density we have
\begin{equation} 
N_{k}=\sum_{n=0}^{\infty }\theta \left( \mu _{k}-n\omega _{k}\right) =1+ \left[ 
\frac{\mu _{k}}{\omega _{k}}\right] \,\,,  \label{3.9m} 
\end{equation}
where $[x]$ denotes the integer part  of the number  $x$. Equation 
(\ref{3.9m}) shows, in particular, that at $\omega _{k}>\mu _{k}$  
we have $N_{k}=1$,  i.e.,
the field structure corresponds to a flat Minkowski space,
with the result that $\omega _{k}< \mu _{k}$  is the range 
of wave vectors in
which nontrivial field properties are expected to show up.

It can easily be verified that from the effective-field
standpoint, the ground state $\Phi _{0}$ is a vacuum state, i.e.,
$a_{k}\Phi _{0}=0$.  On the other hand, the given state can be characterized
by a nonvanishing particle number density. Indeed,
for any wave number we have
\begin{equation} 
\overline{n}_{k}=\sum_{n=0}^{\infty }n\theta \left( \mu _{k}-n\omega 
_{k}\right) =\frac{1}{2}\left( 1+\left[ \frac{\mu _{k}}{\omega _{k}}\right] 
\right) \,\left[ \frac{\mu _{k}}{\omega _{k}}\right] \, 
\end{equation}
with the result that the spectral density of the ground-state
energy is
 \begin{equation} e_{k}=\omega _{k}\overline{n}_{k}=\frac{\omega 
_{k}}{2}\left( 1+\left[ \frac{\mu _{k}}{\omega _{k}}\right] \right) \,\left[ 
\frac{\mu _{k}}{\omega _{k}}\right] .  \label{H0} 
\end{equation}

Since the given particles correspond to the ground state
of the field, in ordinary processes (which do not change the
topology of space) the particles in question are not manifested
explicitly (but they enter into  the renormalization of
the parameters of the observed particles indirectly; here, in
contrast to vacuum fluctuations, the contribution of the particles
is naturally finite). We also note that  although the particles
are bosons, in the ground state they behave like fermions.

One possible explicit manifestation of a residual particle
number density in the ground state is dark matter.  Observations
have shown that dark matter accounts for about 90\% of
visible matter in our universe, and the matter is clearly not of
baryonic origin (see, e.g., Ref. \cite{dm}). Its existence is usually
related to the presence of various hypothetical particles
(Higgs particles, axions, etc.), which for various reasons
cease to interact with ordinary matter. But if this mass is
ascribed to the ground state, then first it becomes obvious
 that the matter is truly dark, and second that the minimum set
incorporating only the particles known at present is 
sufficient.

To describe the properties of dark matter, we begin with
massive bosons ($m\neq 0$).   For the sake of approximation, we
ignore the possible dependence of the parameter $\mu $ on the
wave number $k.$ In this case, to avoid obtaining too large a
value for dark matter, we require that 
\begin{equation} 
\mu ^{2}-m^{2}=z^{2}\ll m^{2}.  
\end{equation} 
Then the ground state contains only one particle per mode in
the wave-number range  $k^{2}\leq z^{2}$,  where $N_{k}=2$. In other
words, massive bosons in the ground state behave like an
ordinary degenerate Fermi gas, and we obtain for the energy
density and particle number density
\begin{eqnarray} 
\varepsilon &=&\frac{1}{L^{3}}\sum_{n,k}n\omega _{k}N\left( k,n\right) =
\frac{g}{2\pi ^{2}}\left( \frac{z^{3}\mu }{4}+\frac{m^{2}}{8}\left( 
z\mu -m^{2}\ln \left( \frac{z+\mu }{m}\right) \right) \right) , \\ 
n &=&\frac{1}{L^{3}}\sum_{n,k}nN\left( k,n\right) =\frac{g}{6\pi ^{2}}z^{3}, 
\end{eqnarray} 
where $g$  is the number of polarization states. In the limit  $z\ll $ $m$,
this expression leads to the well-known nonrelativistic
relationship
\begin{equation} \varepsilon =nm+\frac{3}{2}p,\,\;\;p=\frac{g}{30\pi 
^{2}}\frac{z^{5}}{m}, \label{en} 
\end{equation} 
where $p$ is the gas pressure. The principal contribution to the 
ground-state energy density is provided by the rest mass of
the particle, i.e., in leading order this contribution comes
from dust. Note, however, that the particle pressure is nonzero, 
and it yields a small correction of order  $p/\varepsilon 
\sim z^{2}/m^{2}$
$\sim n^{2/3}/m^{2}$.

We now study particles with zero rest mass (such as 
photons and gravitons). For the ground-state energy density
we have
\begin{equation} \varepsilon =\frac{g}{2\pi 
^{2}}\frac{\mu ^{4}}{4}\xi \left( 3\right) .  \label{d} 
\end{equation} 
The number density of vacuum particles is
\begin{equation} 
n=\frac{g}{2\pi ^{2}}\frac{\mu ^{3}}{3}\xi \left( 2\right) , 
\end{equation} 
where $$\xi \left( s\right) =\sum_{n=1}^{\infty }\frac{1}{n^{s}}.$$ 
The equation of state in this case is ultrarelativistic
($\varepsilon =3p $).

Massless particles are especially intresting, since one
can also measure the intensity of quantum fluctuations of the 
field potentials, which for the ground state (\ref{GST}) are
\begin{equation} 
\left\langle \varphi \left( x\right) \varphi \left( x+r\right) \right\rangle 
=\frac{1}{\left( 2\pi \right) ^{2}}\int\limits_{0}^{\infty 
}\frac{dk}{k}\frac{\sin kr}{kr}\Phi ^{2}\left( k\right) ,  \label{poten} 
\end{equation} 
where
$$\Phi ^{2}\left( k\right) =k^{2}N_{k}=k^{2}\left( 1+\left[ 
\frac{\mu }{k}\right] \right)  . $$  
Thus, at long wavelengths $k\ll \mu $, a substential increase in the level
of quantum fluctuations should be observable in comparison
with pure vacuum noise ($\mu =0$).

\section{Concluding remarks}

We see then that the concept of spacetime foam introduced
by Wheeler should lead to a number of observable
effects in particle theory. The simplest are the emergence of
dark matter and an increase in the intensity of quantum noise 
in the field potentials. In Sec 5 we calculated such effects
under the assumption that the field is in the ground state.
However, the results can easily be generalized to a situation
in which the state of the fields is characterized by nonzero
temperature $T^{\ast }$.   Since processes associated with changes in
the topology of space are the first to stop in the early stages
of the evolution  of the universe, we expect that $T^{\ast }\ll T_{\gamma }$ 
($T_{\gamma }$ is temperature of the microwave background radiation). On 
the  other hand, given the value of  $\mu $ in Eq.  (\ref{d}), we can
obtain an upper bound $\mu ^{\ast }\sim 60T_{\gamma }$ . 
Thus, we expect  $T^{\ast }$ to be much less than $ \mu $, and the
temperature corrections to the   
ground state  (\ref{GST}) to be small. Note, however, that the nature
of the fluctuations of the field potentials in  (\ref{poten})  can change
substantially if the temperature is nonzero \cite{kir94}.

In addition  to the effects studied in this paper, there
clearly remain many phenomena that require additional
investigation. For  example, given the existence of self-action,
the ground state   (\ref{GST}) can be transformed, which can lead to 
the emergence of scalar Higgs fields (by analogy with the
well-known Cooper effect in superconductivity). Such fields
are needed, in turn, to generate particle masses in grand
unification theories. Note that in fields with self-action, a
nonvanishing particle number density in the ground state automatically
leads to the emergence of massive excitations,
although the upper bound on masses that can be derived from
cosmological constraints on the value of  $\mu $ is many orders of
magnitude less than the values observed in particle theory.

Another possibility is that when measuring the Casimir force \cite{Cas48,Sp58},
one must expect an anomalous dependence on distance
at scales exceeding the value of  $1/\mu .$

The author is grateful to D. Turaev for useful discussions
at all stages of the research, and to M. Rainer for invitation 
to Potsdam University,  where a substantial portion of 
this research was done. This work was supported by grants
from Russian Fund for Fundamental Research (Grant No. 
98-02-16273) and DFG (Grant No.  436 RUS 113/236/0(R)).

\end{document}